\documentstyle[12pt,epsfig]{article}
\makeatletter
\def\vereq#1#2{\lower3pt\vbox{\baselineskip1.5pt \lineskip1.5pt
\ialign{$\m@th#1\hfill##\hfil$\crcr#2\crcr\sim\crcr}}}
\makeatother
\begin{document}

\begin{titlepage}
\begin{flushright}
{DPNU-01-35}
\end{flushright}

\vskip 1cm
\begin{center}
{\large\bf Fermion Mass Hierarchy in  Six Dimensional \\ 
$SO(10)$ Grand Unified Theory on a $T^2/Z_2$ Orbifold}
\vskip 1cm {\normalsize N. Haba$^{1,2}$, T. Kondo$^2$ and Y. Shimizu$^2$}
\\
\vskip 0.5cm {\it $^1$Faculty of Engineering, Mie University, Tsu, Mie,
  514-8507, Japan}\\
\vskip 0.5cm {\it $^2$Department of Physics, Nagoya University, Nagoya, 
  464-8602, Japan}
\end{center}
\vskip .5cm

\begin{abstract}

We suggest a simple supersymmetric $SO(10)$ grand unified theory 
 in 6 dimensions which produces the suitable fermion
 mass hierarchies. 
The 5th and 6th dimensional coordinates 
 are compactified on a $T^2/Z_2$ orbifold.
The gauge and Higgs fields propagate in 6 dimensions while
 ordinal chiral matter fields are localized in 4 dimensions.
The orbifolding and boundary conditions realize the gauge 
symmetry reduction, 
 $SU(3)_C\times SU(2)_L \times U(1)_Y \times U(1)_X$, and 
 the triplet-doublet mass splitting. 
We introduce extra three sets of vector-like heavy fields: 
 two sets propagate in 6 dimensions and 
chiral fields which couple to them are defined as the 1st generation, and
 one set propagates 
 in 5 dimensions and chiral fields which couple to them are
 defined as the 2nd generation. 
The suitable fermion mass hierarchies are generated by integrating out
 these vector-like heavy fields.

\end{abstract}
\end{titlepage}
\setcounter{footnote}{0}
\setcounter{page}{1}
\setcounter{section}{0}
\setcounter{subsection}{0}
\setcounter{subsubsection}{0}

\section{Introduction}
\label{sec:intro}
Grand unified theories (GUTs) are appealing models in which the three
gauge groups are unified at a high energy scale. The precise measurements
of the LEP experiments seem to suggest that the three gauge couplings are
unified at about $10^{16}$ GeV with particle contents of the minimal 
supersymmetric (SUSY) standard model (MSSM). 
However, one of the most serious problems to construct a model of GUTs 
is how to realize the mass splitting  between the triplet and 
 the doublet Higgs particles  in the Higgs sector. 
This problem is so-called  triplet-doublet (TD) splitting problem. 
Recently, a new idea for solving the TD splitting problem 
 has been suggested in 5 dimensional (5D) $SU(5)$ 
 GUT where the 5th dimensional 
 coordinate is compactified on an
 $S^1/(Z_2 \times Z_2')$ 
 orbifold \cite{5d}-\cite{ohio},
where only 
 Higgs and gauge fields can propagate in
 5 dimensions. 
The orbifolding makes the $SU(5)$ gauge group
 reduce to the SM gauge group and 
 realizes the TD splitting since  
 the doublet (triplet) Higgs fields have (not)
 Kaluza-Klein zero-modes. 
{}Following this new idea, 
 a lot of works are made progress in the directions of 
 larger unified gauge symmetry and higher 
 space-time dimensions\cite{6d}-\cite{kawamoto}. 
In these models, gauge symmetry and supersymmetry are broken through 
 orbifold projections and/or the Scherk-Schwarz mechanism\cite{SS}.
Especially, the reduction of $SO(10)$ gauge symmetry 
and the TD splitting solution are 
 discussed in 6D theory in Refs.\cite{ABC}\cite{HNOS}. 
The gauge and Higgs fields 
propagate in 6 dimensions, and the suitable
 orbifolding and boundary conditions on the fixed points 
 can induce the gauge symmetry
 reduction, $SO(10) \rightarrow 
 SU(3)_C\times SU(2)_L \times U(1)_Y \times U(1)_X$, 
 and realize the TD splitting by the similar mechanism as 
 the 5D $SU(5)$ GUT on $S^1/(Z_2 \times Z_2')$.
 
As for the trial of inducing the fermion mass hierarchies, 
two scenarios have been proposed so far.
The first scenario has been discussed in the model of
 a 6D $N = 2$ SUSY $SU(5)$ GUT\cite{flavor}, where 
 all matter multiplets including Higgs multiplets 
 are assumed to reside on 4D or 5D 
 space-time, not on 6D 
 space-time. 
In this scenario, 
 the origin of fermion mass hierarchies exists 
 in the volume suppression 
 of wave functions of matter fields. 
However, it is difficult to extend the gauge symmetry to $SO(10)$ and 
put matter fields in 6 dimensions as commented in Ref.\cite{HNOS}. 
The second scenario of inducing 
 the fermion mass hierarchies has been considered 
 in flipped $SU(5)' \times U(1)_X'$ GUT\cite{Fl} in 5 dimensions in Ref.\cite{HSSU}. 
In this model, the suitable fermion mass hierarchies are
 realized, since 
 only fields which correspond to 
 the {\bf 10} representation matter fields 
 of Georgi-Glashow $SU(5)$ GUT\cite{GG} 
 have Kaluza-Klein zero-modes.
However, 
 the small parameter which produces the fermion mass 
 hierarchies has nothing to do with the 
 extra dimension, since it is a free parameter representing 
 the mixing angle between the ordinal 
 chiral generations and
 extra vector-like fields.

In this paper, we would like to 
 suggest more natural mechanism of producing
 fermion mass hierarchies by combining 
 above two scenarios. 
We consider an $N = 1$ SUSY ((1,0)-SUSY) 
 $SO(10)$ GUT in 6 dimensions where
 the 5th and 6th dimensional coordinates are 
 compactified on a $T^2/Z_2$ orbifold\cite{HNOS}. 
The gauge and Higgs fields live in 6 dimensions 
while ordinal chiral matter fields are 
 localized in 4 dimensions.
$N = 1$ SUSY ((1,0)-SUSY) in 6 dimensions
requires the gauginos to have opposite chirality from 
the matter fermions which must all share the same chirality
\cite{others3}\cite{anomaly2}.
This feature strongly constrains our model 
when discussing 6D gauge anomaly. 
The orbifolding and boundary conditions make the $SO(10)$ 
 gauge group be broken to 
 $SU(3)_C\times SU(2)_L \times U(1)_Y \times U(1)_X$ and
 realize the TD splitting. 
In addition to the three-generation chiral matter
 fields, we will introduce extra three sets of 
 vector-like 
 matter fields : 
 two sets of $\mathbf{16}$ and 
 $\overline{\mathbf{16}}$, and four $\mathbf{10}$ 
 representation fields\footnote{ 
 These $\mathbf{10}$ 
 representation matter fields should have 
 the same 6D chirality as $\mathbf{10}$ representation 
 Higgs fields for the 6D gauge anomaly cancellation\cite{others3}.}  
 propagate in 6 dimensions, and 
 chiral fields which couple to $\mathbf{16}$ and 
 $\overline{\mathbf{16}}$ 
 are defined as the 1st generation,
 and one set of $\mathbf{16}$ and 
 $\overline{\mathbf{16}}$ propagates 
 in 5 dimensions and chiral fields which couple to them are
 defined as the 2nd generation. 
We will introduce the $\mathbf{16}$ and $\overline{\mathbf{16}}$ 
 Higgs multiplets on the 4D brane which break the $U(1)_X$ gauge 
 symmetry. 
We assume the GUT scale of 
 $U(1)_X$ breaking, then suitable scale of 
 Majorana masses of right-handed 
 neutrinos are obtained.
The mixing angles between the chiral fields and 
 extra generations will be determined by
 the volume suppression factors. 
The suitable fermion mass hierarchies will be
 generated by integrating out 
 these extra vector-like heavy fields. 
Moreover, the large (small) flavor mixings in the lepton 
 (quark) sector will be naturally explained in order.

In section 2, we show the field contents of 
 this model and discuss an orbifold compactification. 
In section 3, we will discuss $U(1)_X$ symmetry breaking, Majorana
 masses of right-handed neutrinos, and masses of vector-like matter fields.
In section 4, we will see the mechanism of
 generating the fermion mass hierarchies. 
Section 5 gives summary and discussions.


\section{$SO(10)$ GUT on $T^2/Z_2$}
\subsection{$T^2/Z_2$ orbifold}


Let us show the structure of the extra dimensions at first. 
The extra dimensions are compactified on a torus $T^2$, whose
coordinates can be represented by the complex plane, $z \equiv x_5 + ix_6$.
The structure of extra 2D spaces are 
characterized by reflection $P$ and translations $T_i$ ($i = 1, 2$). 
Under the reflection, $z$ is transformed into $-z$, which corresponds
to the $\pi$ rotation on the complex plane. 
The translation symmetry is defined by two vectors $e_i\ (i=1,2)$.
Under the translation, $z$ is transformed into $z + e_i$.
We define ($e_1$, $e_2$) = ($2\pi R_5$, $2\pi R_6$), 
in which $R_5$ and $R_6$ are the two radii of $T^2$.
When the following identifications for the transformations are imposed on 
the 6D scalar field $\Phi(z)$\cite{HNOS} as 
\begin{eqnarray}
\Phi(-z) = P \Phi(z),\\
\Phi(z+e_i) = T_i \Phi(z),
\end{eqnarray}
the scalar field $\Phi(x^\mu,x_5,x_6)$ is divided into 

\begin{eqnarray}
\Phi_{+++}(x^{\mu},x^5,x^6) &=& \frac{1}{\pi\sqrt{2R_5R_6}}
\sum_{m,n=0}^{\infty} \frac{1}{\sqrt{2^{\delta_{m,0}\delta_{n,0}}}} 
\phi_{+++}^{(m,n)}(x^{\mu})
\cos\left(\frac{mx^5}{R_5}+\frac{nx^6}{R_6}\right) , \label{zero}\\
\Phi_{++-}(x^{\mu},x^5,x^6) &=& \frac{1}{\pi\sqrt{2R_5R_6}} \sum_{m,n=0}
^{\infty} \phi_{++-}^{(m,n+\frac{1}{2})}(x^{\mu})
\cos\left(\frac{mx^5}{R_5}+\frac{(n+\frac{1}{2})x^6}{R_6}\right) ,\label{1}\\
\Phi_{+-+}(x^{\mu},x^5,x^6) &=& \frac{1}{\pi\sqrt{2R_5R_6}} \sum_{m,n=0}
^{\infty} \phi_{+-+}^{(m+\frac{1}{2},n)}(x^{\mu})
\cos\left(\frac{(m+\frac{1}{2})x^5}{R_5}+\frac{nx^6}{R_6}\right) ,\\
\Phi_{+--}(x^{\mu},x^5,x^6) &=& \frac{1}{\pi\sqrt{2R_5R_6}} \sum_{m,n=0}
^{\infty} \phi_{+--}^{(m+\frac{1}{2},n+\frac{1}{2})}(x^{\mu})
\cos\left(\frac{(m+\frac{1}{2})x^5}{R_5}+\frac{(n+\frac{1}{2})x^6}{R_6}\right)
 ,~~~\\  
\Phi_{---}(x^{\mu},x^5,x^6) &=& \frac{1}{\pi\sqrt{2R_5R_6}} \sum_{m,n=0}
^{\infty} \phi_{---}^{(m+\frac{1}{2},n+\frac{1}{2})}(x^{\mu})
\sin\left(\frac{(m+\frac{1}{2})x^5}{R_5}+\frac{(n+\frac{1}{2})x^6}{R_6}\right)
 ,~~~\\ 
\Phi_{--+}(x^{\mu},x^5,x^6) &=& \frac{1}{\pi\sqrt{2R_5R_6}} \sum_{m,n=0}
^{\infty} \phi_{--+}^{(m+\frac{1}{2},n)}(x^{\mu})
\sin\left(\frac{(m+\frac{1}{2})x^5}{R_5}+\frac{nx^6}{R_6}\right) ,\\  
\Phi_{-+-}(x^{\mu},x^5,x^6) &=& \frac{1}{\pi\sqrt{2R_5R_6}} \sum_{m,n=0}
^{\infty} \phi_{-+-}^{(m,n+\frac{1}{2})}(x^{\mu})
\sin\left(\frac{mx^5}{R_5}+\frac{(n+\frac{1}{2})x^6}{R_6}\right) ,\\ 
\Phi_{-++}(x^{\mu},x^5,x^6) &=& \frac{1}{\pi\sqrt{2R_5R_6}} \sum_{m,n=0}
^{\infty} \phi_{-++}^{(m,n)}(x^{\mu})
\sin\left(\frac{mx^5}{R_5}+\frac{nx^6}{R_6}\right)\label{2},
\end{eqnarray}
according to the eigenvalues $( \pm, \pm, \pm)$ 
 of the $Z_2$ parity, $T_1$, and $T_2$, respectively. 
The eigenvalues of $P$, $T_1$, and $T_2$ are  $+1$ or $-1$ by definition.
Notice that only $\Phi_{+++}$ in Eq.(\ref{zero}) has a massless zero-mode 
and survives in the low energy. 
Other scalar
 fields in Eqs.(\ref{1})-(\ref{2}) have Kaluza-Klein masses.
The physical space can be taken as  $0 \leq x_5 < 2 \pi R_5$ and 
$0 \leq x_6 \leq \pi R_6$. 
There are four fixed points at 
 $z = 0$, $\pi R_5$, $i\pi R_6$ and $\pi R_5 + i\pi R_6$ on 
the $T^2/Z_2$ orbifold.
We put four 3-branes ($O$-, $O_1$-, $O_2$-, $O_3$-brane) at these
 fixed points as in Fig. \ref{complex plane}. 

Next we show the 6D bulk action of gauge multiplet\cite{6DLag}, which
 is described in terms of 4D $N=1$ vector $(V)$ and 
chiral $(\Phi)$ supermultiplet as 
\begin{eqnarray}
  S &=& \int d^6 x \Biggl\{ {1 \over 4 k g^2} 
    {\rm Tr} \left[ \int d^2\theta {\cal W}^{\alpha} {\cal W}_{\alpha}
    + {\rm h.c.} \right] \nonumber\\
  && + \int d^4\theta {1 \over k g^2} 
    {\rm Tr} \Biggl[ (\sqrt{2} \partial^\dagger + \Phi^\dagger) 
    e^{-V} (-\sqrt{2} \partial + \Phi) e^{V} + 
    \partial^\dagger e^{-V} \partial e^V  \Biggr] \Biggr\},
\label{eq:action}
\end{eqnarray}
where $V=V^a T^a$, $\Phi=\Phi^a T^a$, ${\rm Tr}[T^aT^b]=k\delta^{ab}$ 
and $\partial=\partial_5 - i \partial_6$.
The orbifold $Z_2$ parity shows 
%
\begin{eqnarray}
  V(-z) &=& P V(z) P^{-1},\\
  \Phi(-z) &=& -P \Phi(z) P^{-1},\label{Phi}
\end{eqnarray}
%
since $V$ and $\Phi$ have opposite $Z_2$ parity eigenvalues with each other.
When we take the $Z_2$ parity operator as $P= \sigma_0\otimes I_5$
and impose the $Z_2$ invariance, 
6D $N = 1$ SUSY is broken into 4D $N = 1$ SUSY 
since $\Phi$ in Eq.(\ref{Phi}) vanishes on the 4D 
brane ($z = 0$).
As for the translations, we require the following identifications:
\begin{eqnarray}
  V(z+e_i) &=& T_i V(z) T_i^{-1}, \\
  \Phi(z+e_i) &=& T_i \Phi(z) T_i^{-1}.
\end{eqnarray}
If $T_i$ acts non-trivially on the $SO(10)$ gauge symmetry, 
the $SO(10)$ gauge symmetry is broken into its subgroups
on the fixed points. In this paper we adopt the translations 
as $(T_1, T_2) = (T_{51}, T_{5'1'})$ where 
$T_{51} = \sigma_2\otimes I_5$ and $T_{5'1'} = 
\sigma_2\otimes diag.(1,1,1,-1,-1)$, which commute the generators
of the Georgi-Glashow $SU(5)\times U(1)_X$ and 
the flipped $SU(5)'\times U(1)'_X$ groups, respectively.
As shown in Table \ref{fixed points} and Fig. \ref{complex plane},
the Georgi-Glashow $SU(5)\times U(1)_X$ and 
the flipped $SU(5)'\times U(1)'_X$ symmetries 
remain on $O_1$- and $O_2$-brane, respectively. 
Since $T_1 T_2 = T_{422}\equiv\sigma_0\times diag.(1,1,1,-1,-1)$,
there is the Pati-Salam $SU(4)_C \times SU(2)_L\times SU(2)_R$
symmetry\cite{PS} on $O_3$-brane. 
Note that only an intersection of gauge symmetries on $O_1$- and
$O_2$-branes, which is $SU(3)_C \times SU(2)_L \times U(1)_Y \times U(1)
_X$,
can survive in 4D branes\cite{ABC}\cite{HNOS}. Therefore the
$SO(10)$ gauge symmetry
is broken down to the $SU(3)_C \times SU(2)_L \times U(1)_Y \times
U(1)_X$ gauge symmetries\cite{ABC}\cite{HNOS}.\footnote
{The TD mass splitting and 
the same reduction of the unified gauge
symmetry are realized as long as $T_1 \neq T_2$ 
for $T_i = T_{51}, T_{5'1'}, T_{422}$\cite{HNOS}.
However, in order to realize the fermion mass hierarchies by the
mechanism which we will propose, we 
must choose $T_{51}$ for $T_1$. The following
discussions in this paper are not changed even if we choose 
$(T_1, T_2) = (T_{51}, T_{422})$, in which $O_3$-brane 
has flipped $SU(5)'\times U(1)_X'$
gauge symmetry.}



\begin{table}
\begin{center}
\begin{tabular}{|c|c|} \hline
  $z$     & gauge symmetry 
\\ \hline
  $0$             & $SO(10)$ \\
  $\pi R_5$       & $SU(5) \otimes U(1)_X$ \\
  $i \pi R_6$     & $SU(5)' \otimes U(1)'_X$ \\
  $\pi(R_5+iR_6)$ & $SU(4)_C \otimes SU(2)_L\otimes SU(2)_R$
\\ \hline
\end{tabular}
\end{center}
\caption{Gauge symmetry on each of the four fixed points.}
\label{fixed points} 
\end{table}

\subsection{Higgs and Matter configurations}
\begin{figure}
\begin{center}
\epsfig{file=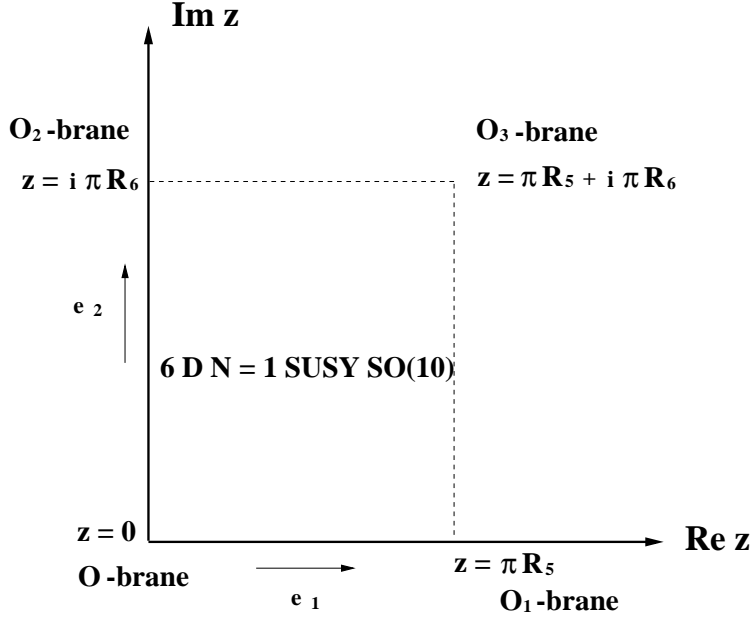,width=10cm}
\caption{The $T^2/Z_2$ orbifold on the complex plane. 
$(z \equiv x_5 +ix_6)$.}
\label{complex plane}
\end{center}
\end{figure}

Let us show the Higgs and matter configurations in 
 our model of 6D $N=1$ SUSY $SO(10)$ GUT.
 First, we discuss 6D gauge anomaly.
Since the 6D theory is the chiral theory, 
 the 6D anomalies must be canceled 
 in addition to 4D anomalies (so-called zero-mode anomalies). 
The irreducible 6D gauge anomalies must be canceled 
 by only the field propagating in 6 dimensions.\footnote
{Other reducible anomalies can be canceled by the Green-Schwarz
mechanism \cite{GS}. For detailed discussions on the anomaly
cancellation, see Refs.\cite{others3}\cite{anomaly}.}
Moreover, 6D $N = 1$ SUSY ((1,0)-SUSY) algebra requires 
the gauginos to have opposite chirality from 
the matter fermions which must all share the same 
 chirality\cite{others3}\cite{anomaly2}.
So, 6D $N = 1$ SUSY ((1,0)-SUSY) determines the relation of 
6D chiralities between the
gauge, Higgs and matter multiplets, automatically.
After this, the 6D chiralities of all matter and Higgs 
fields which we introduced are the same each other, 
and opposite to those of gauginos.
We construct a model based on 
6D $N=1$ SUSY $SO(10)$ GUT.
We know the 6D irreducible gauge anomaly do cancel between
a gauge multiplet and two ${\mathbf{10}}$ hypermultiplet or 
between a ${\mathbf{16}}$ (or  
${\overline{\mathbf{16}}}$) and a ${\mathbf{10}}$ hypermultiplet. 

We consider a situation that the 6D gauge multiplet $(V, \Phi)$ 
and two Higgs multiplets 
$\bf{H_{10}} =  (\rm{H_{10}}, \rm{H^{c}_{\overline{10}}}) $
and  $\bf{H'_{10}} = (\rm{H'_{10}}, \rm{H'^c_{\overline{10}}})$
propagate in the 6D bulk.
The gauge multiplets and $\mathbf{10}$ representation Higgs 
multiplets must have opposite chirality in our model.
These field contents are free from gauge anomaly in 
6 dimensions\cite{others3}\cite{anomaly}.
The assignments of $P$, $T_1$, and $T_2$ are listed in 
Table \ref{tb:particles3}, which shows that only the doublet
Higgs components have $(+,+,+)$ eigenvalues. It means that the doublet
Higgs components have zero-modes while the triplet Higgs components
have Kaluza-Klein masses. This is nothing but the TD splitting
realization\cite{ABC}\cite{HNOS}.

As for matter fields, we assume the ordinal chiral matter fields with 
three generations,
\begin{eqnarray}
\label{miracle}
& & {\bf 16}_i = {\bf 10}_i + {\overline{\bf 5}}_i + {\bf 1}_i, \; \\
& & {\bf 10}_i = ( Q, 
       \overline{U}, \overline{E} )_i \;, 
    {\overline{\bf 5}_i} = ( \overline{D}, L  )_i \;, 
    {\bf 1}_i = ( \overline{N} )_i, \; \nonumber
\end{eqnarray}
are 
localized on the 4D 
brane at $z = 0$ (O-brane) in Fig. \ref{complex plane}, 
where the generation is denoted by $i = 1, 2, 3$. 
The superpotential of the Yukawa sector 
 on the brane at $z = 0$ is given by
\begin{equation}
\label{W_Y}
\label{Yukawa1}
 W_Y = \left(
 \frac{y^u_{ij}}{M_*}{\bf H_{10}} {\bf 16}_i {\bf 16}_j
+ \frac{y^d_{ij}}{M_*}{\bf H'_{10}} {\bf 16}_i {\bf 16}_j
\right)\delta(x_5)\delta(x_6),
\end{equation}
in which $M_*$ is a ultraviolet cut-off scale and
${\bf H}$s represent 6D Higgs fields. 
Here we assume  that there are no hierarchies in Yukawa couplings ${y_{ij}}$s 
in Eq.(\ref{W_Y}), and all elements are the same order of 
magnitude.

In addition to above three-generation chiral matter
 fields, we introduce extra three sets of 
 vector-like matter fields 
: (1) two sets of ${\mathbf 16}$ 
 and ${\overline{\mathbf 16}}$ representation matter fields, 
($\psi_{{\bf 16}_4}$, $\psi_{\overline{{\bf 16}}_4}$, 
$\psi'_{{\bf 16}_4}$, $\psi'_{\overline{{\bf 16}}_4}$), 
 which we call the 4th generation,
 and four ${\bf 10}$ representation
 matter multiplets, $\psi_{\mathbf{10}_{\alpha}}$ 
($\alpha = 1, \cdots ,4$), 
 can propagate in 6 dimensions. 
(2) the other 
 set ($\psi_{{\bf 16}_5}$, $\psi_{\overline{{\bf 16}}_5}$), 
 which we call the 5th generation, can propagate 
 in 5 dimensions. 
The differences between two sets of 
the 4th generation, ($\psi_{{\bf 16}_4}$,
 $\psi_{\overline{{\bf 16}}_4}$) and 
($\psi'_{{\bf 16}_4}$, $\psi'_{\overline{{\bf 16}}_4}$), exist only in 
$T_2$ eigenvalues.
We assume $\psi_{{\bf 16}_5}$ and $\psi_{\overline{{\bf 16}}_5}$ 
reside on $x_6 = 0$ ($SO(10)$ fixed line).
We define the 1st generation chiral fields as those which couple to 
the 4th generation, and the 2nd generation chiral fields 
as those which couple to the 5th generation.
We give $P$, $T_1$ and $T_2$ eigenvalues and the mass spectra 
of the 4th and 5th generations in Table \ref{tb:particles1} 
and Table \ref{tb:particles2}, respectively.
As we show in section \ref{sec:mass}, suitable fermion mass
 hierarchies are generated by integrating out these extra
$\mathbf{16}$ and $\overline{\mathbf{16}}$
heavy fields.\footnote{ 
The extra generation fields, 
 $\psi_{{\bf 16}_4}$,
 $\psi_{\overline{{\bf 16}}_4}$, 
 $\psi'_{{\bf 16}_4}$, $\psi'_{\overline{{\bf 16}}_4}$, 
 $\psi_{{\bf 16}_5}$, and $\psi_{\overline{{\bf 16}}_5}$, 
 also have the Yukawa 
 interactions like Eq.(\ref{W_Y}). 
These interactions are suppressed by the volume suppression factors 
 since extra generation fields propagate in the bulk. 
These Yukawa interactions play the crucial roles 
 for generating the fermion mass hierarchies as will seen 
 in the section 4. }

\section{$U(1)_X$ breaking, $M_R$, and masses of extra generations}

Now let us see 
 the 
 vector-like mass terms generated through 
 $U(1)_X$ symmetry breaking.  
In the previous section we have shown that the $SO(10)$ gauge symmetry
is broken down to the $SU(3)_C \times SU(2)_L \times U(1)_Y \times
U(1)_X$ by the orbifolding and boundary conditions. We must break the $U(1)_X$
symmetry in order to obtain the SM gauge group at the electroweak scale.
Unfortunately the $U(1)_X$ breaking by a simple orbifolding has not been
known so far. 
Thus, here we introduce an additional Higgs multiplets 
($H_{16}$, $H_{\overline{16}}$) 
which are localized on 
$O$-brane. 
We assume these additional Higgs multiplets have the vacuum 
expectation values (VEVs) of $\langle H_{16} \rangle = \langle
H_{\overline{16}} \rangle \equiv v_N$ 
 of order about $10^{16}$ GeV
 in the direction of $B-L$ in a
 scheme.\footnote{
The $U(1)_X$ gauge symmetry breaking 
 might be realized if we introduce the $SU(5)$ singlet fields 
 possessing non-zero $U(1)_X$ charge, its superpotential 
 on the $O_1$-brane 
 ($SU(5) \times U(1)_X$ fixed point), and messenger fields 
 which mediate $U(1)_X$ breaking from $O_1$-brane to 
 $O$-brane\cite{HNOS}.
}
The energy scale of the $U(1)_X$ breaking should be related to 
values of Majorana masses of the right-handed neutrinos. 
$v_N$ of order $10^{16}$ GeV gives 
 the suitable mass scale of the light neutrinos, 
 since the neutrino Yukawa couplings are unified with
 the up-type quark Yukawa couplings as in Eq.(\ref{Yukawa1}). 
As will be shown later, the compactification scale is also taken to
be of order $10^{16}$GeV. So we should consider the value of $v_N$ 
is a little bit smaller than that of the
compactification scale.

We impose Peccei-Quinn(PQ) symmetry\cite{PQ} 
 and its charges
 of fields are shown in Table \ref{table:PQ}. 
The PQ-symmetry 
 distinguishes the Higgs multiplets from
 the matter multiplets.

\begin{table}
\begin{center}
\begin{tabular}{|c|c|} \hline
  matter multiplet     &  PQ charge 
\\ \hline
$\mathrm{H}_{10}$, $\mathrm{H}_{10}$ & $-2$\\
  $H_{16}$, $H_{\overline{16}}$ & $-1$\\ 
${\bf 16}_i$ $\psi_{{\bf 16}_I}$,
$\psi_{\overline{\bf 16}_I}$,
$\psi_{\bf {10}_{\alpha}}$     & 1
\\ \hline
\end{tabular}
\end{center}
\caption{PQ charge for Higgs and matter multiplets.}
\label{table:PQ} 
\end{table}

%
The right-handed neutrino mass terms are generated through the interaction,
\begin{equation}
\label{Yukawa2}
W_{N} = \frac{y^{N}_{ij}}{M_*} H_{\overline{16}}H_{\overline{16}} {\bf 
16}_i {\bf 16}_j. 
\end{equation}
This means the right-handed neutrinos obtain the Majorana masses, 
$M_R = y^{N}\frac{v_N^2}{M_*}$, which are of order $10^{14}$ GeV
with $y^{N} = O(1)$.
Thus, the suitable scale of small 
neutrino mass
of $O(10^{-1})$ eV is obtained through the see-saw mechanism \cite{seesaw}.

Furthermore, we can obtain the vector-like mass terms from $v^N$.
As shown above, we define the 1st (2nd) generation as chiral fields which
couple to 4th (5th) generation.
Then, the mass terms relating to the extra generations are given by
\begin{eqnarray}
W_6 &=& H_{16}H_{\overline{16}} \left\{ 
\frac{y_{44}}{M_*^3}\psi_{{\bf 16}_4}\psi_{\overline{{\bf 16}}_4}
+ \frac{y'_{44}}{M_*^3}\psi'_{{\bf 16}_4}\psi'_{\overline{{\bf 16}}_4}
+ \frac{y_{14}}{M_*^{2}}{\bf 16}_1\psi_{\overline{{\bf 16}}_4}
+ \frac{y'_{14}}{M_*^{2}}{\bf 16}_1\psi'_{\overline{{\bf 16}}_4}
 \right.\nonumber\\
&& \left.
+ \frac{y_{55}}{M_*^2}\psi_{{\bf 16}_5}\psi_{\overline{{\bf 16}}_5}
+ \frac{y_{25}}{M_*^{3/2}}{\bf 16}_2\psi_{\overline{{\bf 16}}_5}\right\}
\delta(x_5)\delta(x_6). 
\end{eqnarray}
Here we assume that the vector-like masses which mix the 4th and the
5th generations are forbidden by the fundamental theory.
After integrating out the 5th and the 6th dimensions, 
we obtain the superpotential,
\begin{eqnarray}
\label{W_4}
W_4 &=& \frac{v_N^2}{M_*} 
\left\{y_{44}\epsilon_1^{4} Q^{(0)}_4 \overline{Q}^{(0)}_4
+ y'_{44}\epsilon_1^{4} \left(
U'^{(0)}_4 \overline{U}'^{(0)}_4 + E'^{(0)}_4 \overline{E}'^{(0)}_4
\right) \right. \nonumber\\
&&\left.
+ y_{14}\epsilon_1^{2} Q_1 \overline{Q}^{(0)}_4
+ y'_{14}\epsilon_1^{2}\left( 
\overline{U_1} U'^{(0)}_4 + \overline{E_1} E'^{(0)}_4
\right)\right. 
\nonumber\\
&&\left.
+ y_{55}\epsilon_2^{2} \left(Q^{(0)}_5 \overline{Q}^{(0)}_5 
+ U^{(0)}_5 \overline{U}^{(0)}_5 + E^{(0)}_5 \overline{E}^{(0)}_5
\right) 
+ y_{25}\epsilon_2 \left(Q_2 \overline{Q}^{(0)}_5 
+ \overline{U_2} U^{(0)}_5 + \overline{E_2} E^{(0)}_5
\right)
\right\},\label{W4a}\nonumber\\
\\
&\simeq&
M \left\{\epsilon_1^{4}\left(Q^{(0)}_4 \overline{Q}^{(0)}_4
+ U'^{(0)}_4 \overline{U}'^{(0)}_4 + E'^{(0)}_4 \overline{E}'^{(0)}_4
\right) + \epsilon_1^{2} \left(Q_1 \overline{Q}^{(0)}_4  
\overline{U_1} U'^{(0)}_4 + \overline{E_1} E'^{(0)}_4
\right)\right.  \nonumber\\
&&\left.
+ \epsilon_2^{2} \left(Q^{(0)}_5 \overline{Q}^{(0)}_5 
+ U^{(0)}_5 \overline{U}^{(0)}_5 + E^{(0)}_5 \overline{E}^{(0)}_5
\right) 
+ \epsilon_2 \left(Q_2 \overline{Q}^{(0)}_5 
+ \overline{U_2} U^{(0)}_5 + \overline{E_2} E^{(0)}_5
\right)
\right\},\label{W4}
\end{eqnarray}
in which $^{(0)}$s represent 
the Kaluza-Klein zero-mode, and 
$\epsilon_i$s are the volume suppression factors,
\begin{eqnarray} 
\epsilon_1 \equiv \sqrt{\frac{1}{2 \pi (R_5 R_6)^{1/2} M_*}}, \;\; \;\;  
\epsilon_2 \equiv \sqrt{\frac{1}{2 \pi R_5 M_*}}.
\label{eps}
\end{eqnarray}
In Eq. (\ref{W4}), we use $M \equiv \frac{v_N^2}{M_*}$ and 
assume that all Yukawa
couplings in Eq.(\ref{W4a}) are of $O(1)$.
The powers of suppression factors are determined whether extra
vector-like fields propagate in 5 dimensions or 6 dimensions.

Finally, we comment on the $\mathbf{10}$ representation 
matter fields.
The $\mathbf{10}$ representation matter fields obtain the mass 
 through the following coupling 
 with $H_{16}$ and $H_{\overline{16}}$ Higgs as 
\begin{eqnarray}
W_6 \sim \frac{1}{M_*^3} H_{16}H_{\overline{16}}  
\psi_{{\bf 10}_{\alpha}}
\psi_{\bf {10}_{\beta}}
\delta(x_5)\delta(x_6). 
\end{eqnarray}
This induces the mass 
 of order $10^{10-11} \mathrm{GeV}$ even to 
 the Kaluza-Klein zero-mode of 
 $\psi_{{\bf 10}_{\alpha}}$. 
Therefore four $\psi_{{\bf 10}_{\alpha}}$s 
 do not survive at the low energy.
Since PQ-symmetry prevents $\psi_{{\bf 10}_{\alpha}}$s from 
 having Yukawa interactions 
 with ordinal chiral three generations, 
 $\psi_{{\bf 10}_{\alpha}}$s have no contributions to 
 the fermion mass matrices discussed in next section.

\section{Fermion Mass Hierarchy}
\label{sec:mass}
The fermion mass hierarchies 
 in the chiral matter fields are generated by 
 integrating out the heavy extra vector-like 
 generations\cite{BB}. 
For a demonstration of this mechanism, here
let us show the doublet-quark $(Q_i)$ sector, for example. 
The mass terms of the doublet-quark sector  
 in Eq.(\ref{W4a}) are given by  
\begin{equation}
 W_M = \sum_{i=1}^2 (M \epsilon_i^{2/i} Q_{i} \overline{Q_{i+3}} + 
                     M \epsilon_i^{4/i} Q_{i+3} \overline{Q_{i+3}})
                     . 
\end{equation}
All these fields represent Kaluza-Klein zero-modes and omit
superscript $(0)$ for simplicity. 
Then the light eigenstate $Q_i^{l}$ and the heavy eigenstate $Q_i^{H}$ 
are given by 
\begin{eqnarray}
\label{Q}
 Q_i^{l} &=&  {\epsilon_i^{4/i} \over \sqrt{(\epsilon_i^
 {4/i})^2 + (\epsilon_i^{2/i})^2}}Q_{i}
         -{\epsilon_i^{2/i} \over \sqrt{(\epsilon_i^{4/i})^2 + 
         (\epsilon_i^{2/i})^2}}Q_{i+3}, \\
 Q_i^{H} &=&  {\epsilon_i^{2/i} \over \sqrt{(\epsilon_i^{4/i})
 ^2 + (\epsilon_i^{2/i})^2}}Q_{i}+
          {\epsilon_i^{4/i} \over
\sqrt{(\epsilon_i^{4/i})^2 + (\epsilon_i^{2/i})^2}}Q_{i+3}. 
\end{eqnarray}
$Q_i^{l}$ becomes the $i$-th generation doublet-quark field at the 
 low energy.
When we set $1/R_5 = 1/R_6 = O(10^{16})$ GeV, Eq.(\ref{eps}) induces 
$\epsilon_i \sim 0.04$ . 
Thus, we can regard $\epsilon_i \simeq\lambda^2$, 
where $\lambda$ is the Cabbibo angle, $\lambda \sim 0.2$. 

The mass hierarchy is generated 
 in the mass matrix of the 
 light eigenstate $Q_i^{l}$ by the small factor $\epsilon_i$ 
($\simeq \lambda^2$). 
The fields $\overline{U_i}$ and $\overline{E_i}$ 
 also receive the same effects as Eq.(\ref{Q}) in 
 the light eigenstates, 
 but $\overline{D_i}$, $L_i$, and
 $\overline{N_i}$ do not receive these effects
 since their extra vector-like generations 
 do not have zero-modes as shown in Tables \ref{tb:particles1} 
and \ref{tb:particles2}.
Then, below the electroweak scale, 
mass matrices of the light eigenstates 
in the up quark sector, down quark sector, 
 and charged lepton sector are 
 given by  
\begin{equation}
 m_u^l \simeq \left(
\begin{array}{ccc}
 \epsilon_1^4 & \epsilon_2 \epsilon_1^2 &  \epsilon_1^2  \\
 \epsilon_1^2 \epsilon_2 & \epsilon_2^2  &  \epsilon_2  \\
 \epsilon_1^2  & \epsilon_2  & 1
\end{array}
\right)  v, \;\;
 m_d^l \simeq \left(
\begin{array}{ccc}
 \epsilon_1^2 & \epsilon_1^2 &  \epsilon_1^2  \\
 \epsilon_2 & \epsilon_2  &  \epsilon_2  \\
  1 & 1 & 1 
\end{array}
\right) \overline{v}, \;\;
 m_e^l \simeq \left(
\begin{array}{ccc}
 \epsilon_1^2 & \epsilon_2 & 1  \\
 \epsilon_1^2 & \epsilon_2  & 1 \\
 \epsilon_1^2 & \epsilon_2 & 1 
\end{array}
\right)  \overline{v}, \;\;
\label{mass}
\end{equation}
respectively, 
where $\overline{v} \equiv \langle h_W \rangle$,   
 $v\equiv \langle h_{\overline{W}}' \rangle$. 
We assume Yukawa couplings of Eq.(\ref{Yukawa1}) 
to be $y \epsilon_{i}^2 = O(1)$. 
Each element of Eq.(\ref{mass}) is understood to be 
multiplied by $O(1)$ coefficient. 
We write the mass matrices in the basis
 that the left-handed fermions are to the left
 and the right-handed fermions are to the right. 
As a result the suitable mass hierarchies are 
realized\cite{BB}-\cite{anarchy}.
{}Moreover, 
 the small (large) flavor mixings 
 in the quark (lepton) sector are naturally 
 obtained. 
%
Since $SO(10)$ relation in Eq.(\ref{Yukawa1}) suggests 
Yukawa couplings of neutrinos are the same order as those of 
up-sector fields, the neutrino Dirac mass matrix is given by 
\begin{equation}
 m_\nu^D \simeq \left(
\begin{array}{ccc}
1 & 1 & 1 \\
1 & 1 & 1 \\
1 & 1 & 1
\end{array}
\right)  v ,
\label{neu}
\end{equation}
because the fields $L_I$, $\overline{L_I}$, 
 $\overline{N_I}$, and $N_I$ do not have 
 the Kaluza-Klein zero-mode. 
Each element of $m_\nu^D$ has $O(1)$ 
 coefficient. 
Neglecting the contributions from the super-heavy 
 Kaluza-Klein masses, 
 Eqs.(\ref{Yukawa2}) and (\ref{neu}) induce 
 the mass matrix of three light neutrinos $m_\nu^{(l)}$ through
the see-saw mechanism as
%
\begin{equation}
\label{26}
 m_\nu^{(l)} \simeq
 {m_\nu^D m_\nu^D{}^T \over M_R} \simeq 
\left(
\begin{array}{ccc}
 1 & 1 & 1 \\
 1 & 1 & 1 \\
 1 & 1 & 1 
\end{array}
\right)  {v^2 \over M_R}.
\end{equation}
%
Since the right-handed neutrino mass $M_R$ is about 
$10^{14}$ GeV in Eq.(\ref{Yukawa2}), 
we can obtain the suitable mass scale ($O(10^{-1})$ eV), for
 the atmospheric neutrino oscillation experiments.
A suitable choice of $O(1)$ coefficients 
 in the mass matrix can derive the suitable 
 flavor mixings consistent with the neutrino
 oscillation experiments\cite{BB}. 
Above fermion mass matrices yield the 
 suitable mass hierarchies of 
 quarks and leptons. 
They also give us a natural explanation 
 why the flavor mixing in the quark sector 
 is small while the flavor mixing in the 
 lepton sector is large\cite{BB}-\cite{anarchy}.

%

\section{Summary and Discussion}
We have suggested 6D $N = 1$ SUSY 
 $SO(10)$ GUT which can produce the suitable 
fermion mass hierarchies. 
The 5th and 6th dimensional coordinates are 
 compactified on a $T^2/Z_2$ orbifold. 
The gauge and Higgs fields live 
 in 6 dimensions while ordinal chiral matter fields are 
 localized in 4 dimensions. 
The orbifolding and boundary conditions make the $SO(10)$ 
 gauge group be broken to 
 $SU(3)_C\times SU(2)_L \times U(1)_Y \times U(1)_X$, and
 realize the TD splitting. 
In addition to the three-generation chiral matter
 fields, we introduce extra three sets of vector-like  
 matter fields: 
 $\psi_{{\bf 16}_4}+ \psi_{\overline{{\bf 16}}_4}$, 
 $\psi'_{{\bf 16}_4}+ \psi'_{\overline{{\bf 16}}_4}$ 
 and four $\psi_{\mathbf{10}_{\alpha}}$s 
 propagate in 6 dimensions and 
 chiral fields which couple to 
 $\psi_{{\bf 16}_4}+ \psi_{\overline{{\bf 16}}_4}$, 
 $\psi'_{{\bf 16}_4}+ \psi'_{\overline{{\bf 16}}_4}$ 
 are defined as the 1st generation, and
 $\psi_{{\bf 16}_5}+ \psi_{\overline{{\bf 16}}_5}$ propagate 
 in 5 dimensions and chiral fields which couple to them are
 defined as the 2nd generation. 
We introduced the $H_{16}$ and $H_{\overline{16}}$ 
 Higgs fields on the 4D brane 
 and assumed the GUT scale 
 $U(1)_X$ breaking.
This energy scale can induce 
 the masses of vector-like fields, and 
 the mixing angles between the chiral fields and 
 extra generations have been determined by
 the volume suppression factors. 
The suitable fermion mass hierarchies are 
 generated by integrating out 
 these extra vector-like heavy fields. 
The large (small) flavor mixings in the lepton 
 (quark) sector are naturally explained in order.

\section*{Acknowledgment}
We would like to thank H. Murayama and Y. Nomura 
 for useful discussions. 
This work is supported in
 part by the Grant-in-Aid for Science
 Research, Ministry of Education, Culture, Sports, Science and 
Technology, of Japan (No.12740146, No.13001292).


\newpage

\begin{table}
\begin{center}
\begin{tabular}[tb]{|l|l|c|c|c|}
 \hline
 6D Higgs fields & 4D Higgs fields 
 & & \\
 $SO(10)$  &  
 $(SU(3)_C, SU(2)_L, U(1)_Y, U(1)_X)$ 
 & $(Z_2, T_1, T_2)$ &
 mass \\
 \hline\hline
 $\rm{H_{10}}$ &   
 $h_C({\bf 3}, {\bf 1}, {\bf -1/3}, {\bf 2})$
 & $(+, +, -)$ & $B$ \\
 &    
 $h_{W}({\bf 1}, {\bf 2}, {\bf 1/2}, {\bf 2})$  
 & $(+, +, +)$ & $A$ \\
 &
 $h_{\overline{C}}({\bf \overline{3}}, {\bf 1}, {\bf 1/3}, {\bf -2})$
 & $(+, -, +)$ & $C$ \\
 &    
 $h_{\overline{W}}({\bf 1}, {\bf 2}, {\bf -1/2}, {\bf -2})$  
 & $(+, -, -)$ & $D$ \\
 \hline
 $\rm{H^{c}_{10}}$&   
 $h_C^{c}({\bf \overline{3}}, {\bf 1}, {\bf 1/3}, {\bf -2})$
 & $(-, +, -)$ & $B$ \\
 &    
 $h_W^{c}({\bf 1}, {\bf 2}, {\bf -1/2}, {\bf -2})$
 & $(-, +, +)$ & $A'$ \\ 
 &
 $h_{\overline{C}}^{c}({\bf 3}, {\bf 1}, {\bf -1/3}, {\bf 2})$
 & $(-, -, +)$ & $C$ \\
 & 
 $h_{\overline{W}}^{c}({\bf 1}, {\bf 2}, {\bf 1/2}, {\bf 2})$
 & $(-, -, -)$ & $D$ \\
 \hline
 $\rm{H'_{10}}$ &   
 $h'_C({\bf 3}, {\bf 1}, {\bf -1/3}, {\bf 2})$
 & $(+, -, +)$ & $C$ \\
 &    
 $h'_W({\bf 1}, {\bf 2}, {\bf 1/2}, {\bf 2})$  
 & $(+, -, -)$ & $D$ \\
 &
 $h'_{\overline C}({\bf \overline{3}}, {\bf 1}, {\bf 1/3}, {\bf -2})$
 & $(+, +, -)$ & $B$ \\
 &    
 $h'_{\overline{W}}({\bf 1}, {\bf 2}, {\bf -1/2}, {\bf -2})$  
 & $(+, +, +)$ & $A$ \\
 \hline
 $\rm{H'^{c}_{10}}$&   
 $h'^{c}_{C}({\bf \overline{3}}, {\bf 1}, {\bf 1/3}, {\bf -2})$
 & $(-, -, +)$ & $C$ \\
 &    
 $h'^{c}_{W}({\bf 1}, {\bf 2}, {\bf -1/2}, {\bf -2})$
 & $(-, -, -)$ & $D$ \\  
 &
 $h'^{c}_{\overline{C}}({\bf 3}, {\bf 1}, {\bf -1/3}, {\bf 2})$
 & $(-, +, -)$ & $B$ \\
 &
 $h'^{c}_{\overline{W}}({\bf 1}, {\bf 2}, {\bf 1/2}, {\bf 2})$
 & $(-, +, +)$ & $A'$ \\
 \hline
\end{tabular}
\end{center}
\caption{The gauge quantum numbers 
 after the compactification,  
 parity eigenvalues of the $(Z_2, T_1, T_2)$ 
 and the mass spectra at the tree level 
 of the Higgs hypermultiplets, ${\bf {H_{10}}}$ and 
${\bf {H'_{10}}}$, are shown.
The Higgs hypermultiplets which propagate in 6 dimensions 
are given as 
${\bf {H_{10}}} = ({\rm H_{10}}, {\rm H^{c}_{\overline{10}}})$ and 
${\bf {H'_{10}}} = ({\rm H'_{10}}, {\rm H'^{c}_{\overline{10}}})$. 
The tree level masses of 
 these multiplets are represented by the five types below. 
$A = \sqrt{\frac{m^2}{R_5^2} + \frac{n^2}{R_6^2}}$, 
$A' = \sqrt{\frac{(m + 1)^2}{R_5^2} + \frac{n^2}{R_6^2}}$ 
or $\sqrt{\frac{m^2}{R_5^2} + \frac{(n + 1)^2}{R_6^2}}$, 
$B = \sqrt{\frac{m^2}{R_5^2} + \frac{\left(n + 1/2\right)^2}{R_6^2}}$, 
$C = \sqrt{\frac{\left(m + 1/2\right)^2}{R_5^2} + \frac{n^2}{R_6^2}}$ 
and $D = \sqrt{\frac{\left(m + 1/2\right)^2}{R_5^2} 
+ \frac{\left(n + 1/2\right)^2}{R_6^2}}$.
The fields with $(+, +, +)$ eigenvalue have zero-modes and survive 
at low energy.}
\label{tb:particles3}
\end{table}


\begin{table}
\begin{center}
\begin{tabular}[tb]{|l|l|c|c|}
\hline
 6D matter fields & 4D matter fields 
  & & \\
 $SO(10)$ & $(SU(3)_C, SU(2)_L, U(1)_Y, U(1)_X)$ 
 &$(Z_2, T_1, T_2)$ &
 mass\\
 \hline\hline
 ${\bf 16}_4/{\bf \overline{16}}_4$ &    
 $Q_4({\bf 3}, {\bf 2}, {\bf 1/6}, {\bf -1})
/\overline{Q}_4({\bf \overline{3}}, {\bf 2}, {\bf -1/6}, {\bf 1})$
  & $(+, +, +)$ & $A$ \\
 &    
 $\overline{U}_4({\bf \overline{3}}, {\bf 1}, {\bf -2/3}, {\bf -1})$
 /$U_4({\bf 3}, {\bf 1}, {\bf 2/3}, {\bf 1})$
  &  $(+, +, -)$ & $B$ \\
 &
 $\overline{E}_4({\bf 1}, {\bf 1}, {\bf 1}, {\bf -1})$
 /$E_4({\bf 1}, {\bf 1}, {\bf -1}, {\bf 1})$
 &$(+, +, -) $ &$B$ \\
 &    
 $\overline{D}_4({\bf \overline{3}}, {\bf 1}, {\bf 1/3}, {\bf 3})$
 /$D_4({\bf 3}, {\bf 1}, {\bf -1/3}, {\bf -3})$
 &  $(+, -, +)$ & $C$ \\
 &
 $\overline{N}_4({\bf 1}, {\bf 1}, {\bf 0}, {\bf -5})$
 /$N_4({\bf 1}, {\bf 1}, {\bf 0}, {\bf 5})$
 &  $(+, -, +) $& $C$ \\
 &    
 $L_4({\bf 1}, {\bf 2}, {\bf -1/2}, {\bf 3})$
 /$\overline{L}_4({\bf 1}, {\bf 2}, {\bf 1/2}, {\bf -3})$
 &  $(+, -, -)$ & $D$ \\
 
 \hline
 ${\bf \overline{16}}_4^c/{\bf 16}_4^c$ &
  $\overline{Q_4}^{c}({\bf 3}, {\bf 2}, {\bf 1/6}, {\bf -1})$
 /$Q_4^{c}({\bf \overline{3}}, {\bf 2}, {\bf -1/6}, {\bf 1})$
 & $(-, +, +)$ & $A'$ \\
 &    
 $U_4^{c}({\bf \overline{3}}, {\bf 1}, {\bf -2/3}, {\bf -1})$
 /$\overline{U}_4^{c}({\bf 3}, {\bf 1}, {\bf 2/3}, {\bf 1})$
 &  $(-, +, -)$ & $B$ \\
 & 
 $E_4^{c}({\bf 1}, {\bf 1}, {\bf 1}, {\bf -1})$
 /$\overline{E}_4^{c}({\bf 1}, {\bf 1}, {\bf -1}, {\bf 1})$
 &  $(-, +, -)$ & $B$ \\
 &    
 $D_4^{c}({\bf \overline{3}}, {\bf 1}, {\bf 1/3}, {\bf 3})$, 
 /$\overline{D}_4^{c}({\bf 3}, {\bf 1}, {\bf -1/3}, {\bf -3})$
 &  $(-, -, +)$ & $C$ \\
 &    
 $N_4^{c}({\bf 1}, {\bf 1}, {\bf 0}, {\bf -5})$
 /$\overline{N}_4^{c}({\bf 1}, {\bf 1}, {\bf 0}, {\bf 5})$
 &  $(-, -, +) $ & $C$ \\
 &    
 $\overline{L}_4^{c}({\bf 1}, {\bf 2}, {\bf -1/2}, {\bf 3})$
 /$L_4^{c}({\bf 1}, {\bf 2}, {\bf 1/2}, {\bf -3})$
 &  $(-, -, -)$ & $D$ \\
 \hline
 ${\bf 16'}_4/{\bf \overline{16}}'_4$ &    
 $Q'_4({\bf 3}, {\bf 2}, {\bf 1/6}, {\bf -1})
 /\overline{Q}'_4({\bf \overline{3}}, {\bf 2}, {\bf -1/6}, {\bf 1})$
  & $(+, +, -)$ & $B$ \\
 &    
 $\overline{U}'_4({\bf \overline{3}}, {\bf 1}, {\bf -2/3}, {\bf -1})$
 /$U'_4({\bf 3}, {\bf 1}, {\bf 2/3}, {\bf 1})$
 &  $(+, +, +)$ & $A$ \\
 &    
 $\overline{E}'_4({\bf 1}, {\bf 1}, {\bf 1}, {\bf -1})$
 /$E'_4({\bf 1}, {\bf 1}, {\bf -1}, {\bf 1})$,
 &  $(+, +, +)$ & $A$ \\
 &    
 $\overline{D}'_4({\bf \overline{3}}, {\bf 1}, {\bf 1/3}, {\bf 3})$
 /$D'_4({\bf 3}, {\bf 1}, {\bf -1/3}, {\bf -3})$
 &  $(+, -, -)$ & $D$ \\
 &    
 $\overline{N}'_4({\bf 1}, {\bf 1}, {\bf 0}, {\bf -5})$
 /$N'_4({\bf 1}, {\bf 1}, {\bf 0}, {\bf 5})$
 & $(+, -, -) $ & $D$ \\
 &    
 $L'_4({\bf 1}, {\bf 2}, {\bf -1/2}, {\bf 3})$
 /$\overline{L}'_4({\bf 1}, {\bf 2}, {\bf 1/2}, {\bf -3})$
 &  $(+, -, +)$ & $C$ \\
 
 \hline
 ${\bf \overline{16}}'^{c}_4/{\bf 16}'^{c}_4$ &
 $\overline{Q}'^{c}_4({\bf 3}, {\bf 2}, {\bf 1/6}, {\bf -1})$
 /$Q'^{c}_4({\bf \overline{3}}, {\bf 2}, {\bf -1/6}, {\bf 1})$
  & $(-, +, -)$ & $B$ \\
 &    
 $U'^{c}_4({\bf \overline{3}}, {\bf 1}, {\bf -2/3}, {\bf -1})$ 
 /$\overline{U}'^{c}_4({\bf 3}, {\bf 1}, {\bf 2/3}, {\bf 1})$
 &  $(-, +, +)$ & $A'$ \\
 &    
 $E'^{c}_4({\bf 1}, {\bf 1}, {\bf 1}, {\bf -1})$
 /$\overline{E}'^{c}_4({\bf 1}, {\bf 1}, {\bf -1}, {\bf 1})$
 &  $(-, +, +)$ & $A'$ \\
 &    
 $D'^{c}_4({\bf \overline{3}}, {\bf 1}, {\bf 1/3}, {\bf 3})$ 
 /$\overline{D}'^{c}_4({\bf 3}, {\bf 1}, {\bf -1/3}, {\bf -3})$
 &  $(-, -, -)$ & $D$ \\
 &    
 $N'^{c}_4({\bf 1}, {\bf 1}, {\bf 0}, {\bf -5})$
 /$\overline{N}'^{c}_4({\bf 1}, {\bf 1}, {\bf 0}, {\bf 5})$
 &  $(-, -, -) $ & $D$ \\
 &    
 $\overline{L}'^{c}_4({\bf 1}, {\bf 2}, {\bf -1/2}, {\bf 3})$
 /$L'^{c}_4({\bf 1}, {\bf 2}, {\bf 1/2}, {\bf -3})$
 &  $(-, -, +)$ & $C$ \\
 \hline
\end{tabular}
\end{center}
\caption{The gauge quantum numbers after the compactification, 
 parity eigenvalues of the $(Z_2, T_1, T_2)$, 
 and the mass spectra at the tree level of the two sets of 
 vector-like extra matter fields, $\psi_{{\bf 16}_4}$, 
$\psi_{\overline{{\bf 16}}_4}$, $\psi'_{{\bf 16}_4}$ and 
$\psi'_{\overline{\bf 16}_4}$,
 are shown. 
 These vector-like multiplets which propagate in 6 dimensions 
 are given as  $\psi_{{\bf 16}_4} = ({\bf
 16}_4, {\bf \overline{16}}^c_4)$, $\psi_{\overline{{\bf 16}}_4} =
 ({\bf \overline{16}}_4, {\bf {16}}^c_4)$, 
$\psi'_{{\bf 16}_4} = ({\bf 16}'_4, {
 \overline{\bf 16}}'^c_4)$ and $\psi'_{\overline{\bf 16}_4} =
 ({\overline{\bf 16}}'_4, {\bf {16}}'^c_4)$. The tree level masses of 
 these multiplets are represented by the five types below. 
$A = \sqrt{\frac{m^2}{R_5^2} + \frac{n^2}{R_6^2}}$, 
$A' = \sqrt{\frac{(m + 1)^2}{R_5^2} + \frac{n^2}{R_6^2}}$ 
or $\sqrt{\frac{m^2}{R_5^2} + \frac{(n + 1)^2}{R_6^2}}$, 
$B = \sqrt{\frac{m^2}{R_5^2} + \frac{\left(n + 1/2\right)^2}{R_6^2}}$, 
$C = \sqrt{\frac{\left(m + 1/2\right)^2}{R_5^2} + \frac{n^2}{R_6^2}}$ 
and $D = \sqrt{\frac{\left(m + 1/2\right)^2}{R_5^2} 
+ \frac{\left(n + 1/2\right)^2}{R_6^2}}$. 
The fields with $(+, +, +)$ eigenvalue have zero-modes and survive 
at low energy.
} 
\label{tb:particles1}
\end{table}

\begin{table}
\begin{center}
 \begin{tabular}[tb]{|l|l|c|c|}

\hline
 5D matter& 4D matter fields 
 & &  \\
 fields & 
 & &  \\
 $SO(10)$ & $(SU(3)_C, SU(2)_L, U(1)_Y, U(1)_X)$ 
 &$(Z_2, T_1)$ & 
 mass \\
 \hline\hline
 ${\bf 16}_5$ &    
 $ Q_5({\bf 3}, {\bf 2}, {\bf 1/6}, {\bf -1})$, 
 $\overline{U}_5({\bf \overline{3}}, {\bf 1}, {\bf -2/3}, {\bf -1})$,
 $\overline{E}_5({\bf 1}, {\bf 1}, {\bf 1}, {\bf -1})$
 & $(+, +)$ & $n/R_5$ \\
 &    
 $L_5({\bf 1}, {\bf 2}, {\bf -1/2}, {\bf 3})$,
 $\overline{D}_5({\bf \overline{3}}, {\bf 1}, {\bf 1/3}, {\bf 3})$, 
 $\overline{N}_5({\bf 1}, {\bf 1}, {\bf 0}, {\bf -5})$
 & $(+, -)$ & $(n + \frac{1}{2})/R_5 $ \\
\hline
 ${\bf \overline{16}}_5^{c} $ &  
 $\overline{Q_5}^{c}({\bf 3}, {\bf 2}, {\bf 1/6}, {\bf -1})$,
 $U_5^{c}({\bf \overline{3}}, {\bf 1}, {\bf -2/3}, {\bf -1})$,
 $E_5^{c}({\bf 1}, {\bf 1}, {\bf 1}, {\bf -1})$ 
 & $(-, +)$ & $(n + 1)/R_5$\\
 &   
 $\overline{L}_5^{c}({\bf 1}, {\bf 2}, {\bf -1/2}, {\bf 3})$, 
 $D_5^{c}({\bf \overline{3}}, {\bf 1}, {\bf 1/3}, {\bf 3})$,
 $N_5^{c}({\bf 1}, {\bf 1}, {\bf 0}, {\bf -5})$ 
 & $(-, -)$ &  $(n + \frac{1}{2})/R_5$ \\ 
 \hline
 ${\bf \overline{16}}_5$ &   
 $\overline{Q}_5({\bf \overline{3}}, {\bf 2}, {\bf -1/6}, {\bf 1})$,
 $U_5({\bf 3}, {\bf 1}, {\bf 2/3}, {\bf 1})$,
 $E_5({\bf 1}, {\bf 1}, {\bf -1}, {\bf 1})$
 & $(+, +)$ & $n/R_5$ \\
 &   
 $\overline{L}_5({\bf 1}, {\bf 2}, {\bf 1/2}, {\bf -3})$,
 $D_5({\bf 3}, {\bf 1}, {\bf -1/3}, {\bf -3})$, 
 $N_5({\bf 1}, {\bf 1}, {\bf 0}, {\bf 5})$
 & $(+, -)$ & $(n + \frac{1}{2})/R_5$ \\
 \hline
 ${\bf 16}_5^{c}$ &
 $Q_5^{c}({\bf \overline{3}}, {\bf 2}, {\bf -1/6}, {\bf 1})$, 
 $\overline{U}_5^{c}({\bf 3}, {\bf 1}, {\bf 2/3}, {\bf 1})$,
 $\overline{E}_4^{c}({\bf 1}, {\bf 1}, {\bf -1}, {\bf 1})$
 & $(-, +)$ & $(n + 1)/R_5$ \\
 &  
 $L_5^{c}({\bf 1}, {\bf 2}, {\bf 1/2}, {\bf -3})$,
 $\overline{D}_5^{c}({\bf 3}, {\bf 1}, {\bf -1/3}, {\bf -3})$,
 $\overline{N}_5^{c}({\bf 1}, {\bf 1}, {\bf 0}, {\bf 5})$
 & $(-, -)$ & $(n + \frac{1}{2})/R_5$ \\
 \hline
 \end{tabular}
\end{center}
\caption{The gauge quantum numbers after the compactification,  
 parity eigenvalues of the $(Z_2, T_1)$, 
 and the mass spectra at the tree level of the one set of 
 vector-like extra matter fields, $\psi_{{\bf 16}_5}$ and 
$\psi_{\overline{{\bf 16}}_5}$, are shown. 
These vector-like multiplets are given as 
$\psi_{{\bf 16}_5} = ({\bf 16}_5, {\bf
 \overline{16}}^c_5)$ and $\psi_{\overline{{\bf 16}}_5} =
 ({\bf \overline{16}}_5, {\bf {16}}^c_5)$. 
 The fields with $(+, +)$ eigenvalue have zero-modes and survive 
at low energy.} 
\label{tb:particles2}
\end{table}

%

\begin{thebibliography}{99}
%
%
%
%
%
%
%
%
%
%
%
%
%
%
%
%
%
%


\bibitem{5d}
Y.~Kawamura,
Prog.\ Theor.\ Phys.\  {\bf 103} (2000),  613; ibid 
\  {\bf 105} (2001), 691; ibid {\bf 105} (2001), 999;\\
G.~Altarelli and F.~Feruglio,
Phys.\ Lett.\ B {\bf 511} (2001), 257;\\
A.~B.~Kobakhidze,
Phys.\ Lett.\ B {\bf 514} (2001), 131;\\
L.~J.~Hall and Y.~Nomura,
Phys.\ Rev. D {\bf 64} (2001), 055003;\\
Y.~Nomura, D.~Smith and N.~Weiner,
Nucl.\ Phys.\ B {\bf 613} (2001), 147;\\
A.~Hebecker and J.~March-Russell,
Nucl. Phys. B {\bf 613} (2001), 3;\\
R.~Barbieri, L.~J.~Hall and Y.~Nomura,
hep-ph/0106190;\\
R.~Barbieri, L.~J.~Hall and Y.~Nomura,
hep-th/0107004;\\
A.~E.~Faraggi,
Phys.\ Lett.\ B {\bf 520} (2001), 337;\\
L.~J.~Hall, H.~Murayama and Y.~Nomura,
hep-th/0107245; \\
N. Haba, T.~Kondo, Y.~Shimizu, T.~Suzuki and K.~Ukai,
hep-ph/0108003;\\
Y.~Nomura,
hep-ph/0108170;\\
L.~J.~Hall and Y.~Nomura,
hep-ph/0111068. 

\bibitem{others3}
A.~Hebecker and J.~March-Russell,
hep-ph/0107039.


\bibitem{HSSU}
N. Haba, Y.~Shimizu, T.~Suzuki and K.~Ukai,
hep-ph/0107190. 


\bibitem{ohio}
R.~Dermisek and  A.~Mafi,
hep-ph/0108139.




\bibitem{6d}
T.~Li, 
hep-ph/0108120; Phys.\ Lett.\ B {\bf 520} (2001), 377;\\
L.~J.~Hall, Y.~Nomura and D.~Smith,
hep-ph/0107331;\\
T.~Watari and T.~Yanagida,
Phys.\ Lett.\ B {\bf 519} (2001), 164.


\bibitem{ABC}
T.~Asaka, W.~Buchm\"uller and  L.~Covi,
hep-ph/0108021.

\bibitem{HNOS}
L.~J.~Hall, Y.~Nomura, T.~Okui and  D.~Smith,
hep-ph/0108071.

\bibitem{flavor}
L.~J.~Hall, J.~March-Russell, T.~Okui and D.~Smith,
hep-ph/0108161.

\bibitem{6DLag}
N.~Arkani-Hamed, T.~Gregoire and J.~Wacker,
hep-th/0101233. 

 




\bibitem{kawamoto}
T.~Kawamoto and Y.~Kawamura,
hep-ph/0106163.


\bibitem{SS}
J.~Scherk and J.~H.~Schwarz,
Phys.\ Lett.\ B {\bf 82} (1979), 60; Nucl.\ phys.\ B {\bf 153} (1979),
61.


\bibitem{Fl}
S.~M.~Barr,
Phys.\ Lett.\ B {\bf 112} (1982), 219; 
Phys.\ Rev.\ D {\bf 40} (1989), 2457;\\ 
J.~P.~Derendinger, J.~E.~Kim and D.~V.~Nanopoulos,
Phys.\ Lett.\ B {\bf 139} (1984), 170;\\
I.~Antoniadis, J.~Ellis, J.~S.~Hagelin and D.~V.~Nanopoulos,
Phys.\ Lett.\ B {\bf 194} (1987), 231;\\ 
J.~L.~Lopez and D.~V.~Nanopoulos,
hep-ph/9511266.





\bibitem{GG}
H.~Georgi and S.~L.~Glashow,
Phys.\ Rev.\ Lett.\  {\bf 32}, (1974), 438.





\bibitem{anomaly2}
A.~Sagnotti,
Phys.\ Lett.\ B {\bf 294} (1992), 196.






\bibitem{PS}
J.~C.~Pati and A.~Salam,
Phys.\ Rev.\ D {\bf 10}, (1974), 275.











\bibitem{GS}
M.~B.~Green and J.~H.~Schwarz,
Phys.\ Lett.\ B {\bf 149} (1984), 117.

\bibitem{anomaly}
N.~Borghini, Y.~Gouverneur and M.~Tytgat,
hep-ph/0108094.








\bibitem{PQ}
R.~D.~Peccei and H.~R.~Quinn,
Phys.\ Rev.\ Lett {\bf 38} (1977), 1440;
Phys.\ Rev.\ D {\bf 16} (1977), 1791.
\bibitem{seesaw}
T.\ Yanagida, in {\sl Proceedings of the Workshop on
    Unified Theory and Baryon Number of the Universe}, eds\@. O.\
  Sawada and A.\ Sugamoto (KEK, 1979) p.95; M.\ Gell-Mann, P.\ Ramond
  and R.\ Slansky, in {\sl Supergravity}, eds\@. P.\ van Nieuwenhuizen 
  and D.Z.\ Freedman (North Holland, Amsterdam, 1979) p.315.

\bibitem{BB}
K.~S.~Babu and S.~M.~Barr,
Phys.\ Lett.\ B {\bf 381} (1996), 202;\\
S.~M.~Barr,
Phys.\ Rev.\ D {\bf 55} (1997), 1659.


\bibitem{BB1}
M. J. Strassler, Phys.\ Lett.\ B {\bf 376} (1996), 119;\\
N.~Haba,
Phys.\ Rev.\ D {\bf 59} (1999), 035011;\\
K.~Yoshioka,
Mod.\ Phys.\ Lett.\ A {\bf 15} (2000), 29;\\
J.~Hisano, K.~Kurosawa, and Y.~Nomura, 
Nucl.\ Phys.\ B {\bf 584} (2000), 3;\\
T.~Blazek and S.F.~King, 
Phys.\ Lett.\ B {\bf 518} (2001), 109.





\bibitem{anarchy}
N.~Haba and H.~Murayama,
Phys.\ Rev.\ D {\bf 63} (2001), 053010.










%
%
%





\end{thebibliography}
\end{document}